# Grand Canonical Monte Carlo Simulation of Hydrogen Adsorption in Different Carbon Nano Structures

**Tengfei Luo**
2555 Engineering Building
Mechanical Engineering
Michigan State University
East Lansing, MI 48824
Email: luotengf@msu.edu

**John R. Lloyd**
ASFC 1.316.C RRMC
Rapid Response Manufacturing Center
The University of Texas Pan American
Edinburg, TX 78539
Email: lloyd@egr.msu.edu

**Abstract**
Grand Canonical Monte Carlo (GCMC) simulations are performed to study hydrogen physisorption in different nano carbon porous materials made up of different substructures including carbon nanotubes (CNT), graphene sheets and C60. Hydrogen weight percentage (wt%) at different temperatures with pressure ranging from 1 to 20MPa are predicted. Fugacity and quantum effects on hydrogen adsorption are investigated. Different structural dimensions including the sizes of the substructures and spacing between the substructures are used to study the geometrical effects on hydrogen storage capacity in carbon materials. The calculated results generally agree well with available data from other calculations. It is concluded that CNT arrays, graphite nanofibers (GNF) and C60 intercalated graphite (CIG) are not promising to reach the DOE 6.5 wt% target at room temperature. It is also found that the quantum effect is significant in low temperature hydrogen adsorption and different treatments to account for the quantum effect also influence the predicted wt% differently.

## 1. Introduction

Hydrogen has been recognized as an ideal energy carrier since it is clean, renewable and sustainable. It is a promising alternative fuel to the fossil fuels. However, its implementation in transportation is largely limited by the available storage method. The DOE (U.S. Department of Energy) set a target of 6.5 wt% of hydrogen storage for commercial applications for the year of 2010. However, the goal has not been met. Among various possible storage techniques, physisorption using nanoporous carbon structures has drawn much attention. There are many studies on hydrogen adsorption using single-walled carbon nanotubes (CNT) both experimentally [1-5] and theoretically [6-13, 23, 31]. Other carbon based nano structures such as graphite nanofibers (GNF) [14,15,22], C60 intercalated graphite (CIG) [16], pillared graphene [17] and carbon nanoscrolls [18,19] were also studied mainly by theoretical calculations and simulations. CNT arrays were regarded as a promising material for hydrogen storage since high wt% results were reported experimentally. Dillon et. al. found a hydrogen wt% of 5-10% [3] at ambient pressure near room temperature and Chen et. al [1] reported a 14-20 wt% in Li- or P-doped carbon nanotubes at room temperature. However, some of the experimental results are controversial [20,21] and have not been reproduced due the difficulty of controlling the complicated experimental conditions. On the other hand, many molecular modeling has predicted a much lower wt% (around 1%) of hydrogen storage in various CNT arrays [8,10,11,13,23,31] at room temperatures. Other pure carbon materials also cannot reach the 6.5 wt% target at room temperatures according to various simulation and theoretical perditions [16-19].

In theoretical studies of hydrogen storage in carbon structures, molecular dynamics (MD) [18,24,25], grand canonical Monte Carlo (GCMC) simulation [6,8,10-13,17,19,22,23,31] and first principle calculation [7,9,16,17,19,26] are the methods often used. MD is a good method to investigate the hydrogen diffusion processes. However, MD simulations usually need to simulate a large number of hydrogen molecules to model the reservoir environment and it cannot simulate diffusion processes which may take several minutes to even hours in physical time. GCMC is a suitable approach to simulate the equilibrium state of hydrogen physisorption when the carbon structures are attached to a hydrogen reservoir with well-defined pressure and temperature. First principle calculation is capable of studying possible chemisorption but it is not suitable for large scale simulation.

Although there have been many studies on hydrogen storage in carbon structures, a comprehensive investigation of the factors that influence the hydrogen physisorption wt% in these material is still of interest and such studies will be valuable for further research to improve the wt%. In this work, GCMC simulations are performed on hydrogen adsorption in CNT arrays, GNF and CIG. Fugacity effect and quantum effect are studied. Temperature, pressure and structural dimension effects are also investigated.

## 2. Theory and Model
### 2.1 Monte Carlo method
For a gas adsorption process, the adsorbent structure is attached to a gas reservoir and the gas will diffuse into the adsorbent. At the equilibrium state, the temperature and chemical potential of the adsorbed gas are equal to those of the gas in the reservoir. In a GCMC simulation, the temperature, chemical potential of the reservoir and the adsorbent structures are input as known parameters and the equilibrium number of adsorbed gas molecules can be calculated. For hydrogen adsorption in carbon structures, the adsorbed hydrogen gas is in equilibrium with the hydrogen in the reservoir when the temperature and the chemical potential of the hydrogen inside the carbon structures are equal to those of the reservoir. Further details of GCMC method is well documented in ref. [27].

In a GCMC simulation, instead of setting the chemical potential, it is more intuitive to set the reservoir pressure which is related to the chemical potential by eq. 1 under the ideal gas assumption.

$$\mu = \frac{1}{\beta}\ln(\Lambda^3 \beta P) \qquad (1)$$

where $\mu$ is the chemical potential, $\frac{1}{\beta} = k_B T$, $\Lambda$ is the thermal deBroglie wavelength and $P$ is the reservoir pressure.

**2.2 Potential Model**

In this work, the interaction between hydrogen molecules and the interaction between hydrogen molecules and carbon atoms are modeled by the Lennard-Jones (LJ) potential:

$$U_{LJ}(r) = 4\varepsilon\left[\left(\frac{\sigma}{r}\right)^{12} - \left(\frac{\sigma}{r}\right)^{6}\right] \quad (2)$$

The diatomic hydrogen molecule is treated as an entity. The potential parameters are $\varepsilon_{H_2-H_2} = 0.00316 eV$ and $\sigma_{H_2-H_2} = 2.958 \overset{o}{A}$ for interactions between hydrogen molecules and $\varepsilon_{H_2-C} = 0.0028 eV$ and $\sigma_{H_2-C} = 3.179 \overset{o}{A}$ for interactions between hydrogen molecules and carbon atoms. Such a model has been widely used in GCMC simulations of hydrogen adsorption in carbon structures [6,8,12,13,17]. Beside the aforementioned potential functions, ref. [6,8] also included a quadrupole-quadrupole interaction which was treated by the Coulomb interactions of effective charges. We tested the model which included the quadrupole-quadrupole interaction by simulating hydrogen adsorption in an empty box. The difference of the results from such a model and the model without quadrupole-quadrupole interaction are within 2%. Moreover, since the carbon structures in this work are not charged, there is no Coulomb interaction between hydrogen molecules and carbon atoms. As a result, we ignored the quadrupole-quadrupole interaction in our simulation. The tail correction is applied to our calculation to compensate the cutoff influence.

In our simulations, $1.5 \times 10^7$ operations are performed for equilibration and another $2.0 \times 10^7$ operations are performed for production. Three types of operations (molecule displacement, creation and deletion) with equal probabilities are performed randomly in the GCMC simulation. Periodic boundary conditions (PBC) are applied for all three spatial directions in all cases. Different simulation supercells are used for different carbon structures and these are specified separately in each subsection of section 3.

**3. Simulation and Result**

**3.1 Fugacity and Quantum Effect**

At very low densities, a gas system can be regarded as a system of non-interacting point particles which is called an ideal gas. However, when the pressure increases or the temperature lowers, gas molecules get closer and start interacting with each other. In these conditions, ideal gas model fails. As a result, the fugacity which works as the "corrected pressure" for real gases should be used to describe the reservoir gas, and thus eq. (1) becomes eq. (3) which includes the fugacity correction.

$$\mu = \frac{1}{\beta}\ln(\Lambda^3 \beta \phi P) \quad (3)$$

where $\phi \equiv f/P$ is the fugacity coefficient, $f$ is fugacity of non-ideal gas and $P$ is the ideal gas pressure. At temperature above $0^o c$ and pressures lower than $300 MPa$, the fugacity coefficient can be calculated by the following empirical equation [28]:

$$\ln \phi = c_1 P - c_2 P^2 + c_3 \left[\exp(-P/300) - 1\right]$$
$$c_1 = \exp(-3.8402 T^{1/8} + 0.5410)$$
$$c_2 = \exp(-0.1263 T^{1/2} - 15.980) \quad (4)$$
$$c_3 = 300 \exp(-0.11901 T - 5.941)$$

To test the effect of the fugacity on describing the reservoir hydrogen, hydrogen adsorption in an empty box is simulated, the equilibrium pressure of the box is calculated and plotted versus the set pressure of the reservoir (see Figure 1). With the fugacity coefficient applied, the calculated equilibrium pressure of the empty box becomes closer to the set pressure of the reservoir. This means that the reservoir gas behaves more like interacting LJ gas when fugacity effect is considered.

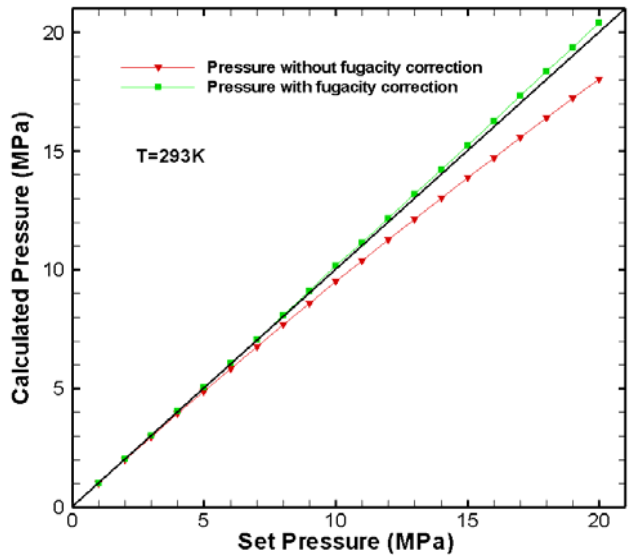

Figure 1. Effect of fugacity on pressure at 293K (the black solid line refers to x=y(x)).

However, there is no empirical fugacity coefficient functions available for hydrogen at temperature lower than $0^o c$. A method is used to fit fugacity coefficients as functions of the set pressures for lower temperatures and it is described later in this section.

Because the hydrogen molecular mass is so small, the quantum effect should be considered as it could contribute a lot to the atomic interactions especially at low temperatures. In this work, we employed the method described in ref. [8] to use a Feynman-Hibbs effective potential to estimate the quantum effects to the order of $\hbar^2$ as found in eq. (5).

$$U_{LJ}^{F-H}(r) = U_{LJ}(r) + \frac{\theta(T)\sigma^2}{24}\left(\frac{d^2 U_{LJ}(r)}{dr^2} + \frac{2}{r}\frac{dU_{LJ}(r)}{dr}\right) \quad (5)$$

where the dimensionless coefficient $\theta(T)$ is defined as

$$\theta(T) = \frac{\hbar^2}{m_r \sigma^2 k_B T}. \quad (6)$$

$\sigma$ refers to the length scale parameter in the LJ potential, $\hbar$ is the reduced Plank constant and $m_r$ is the reduced mass of the LJ pairs ($m_r = 1.6744 \times 10^{-27} kg$ for $H_2 - H_2$ pairs and $m_r = 2.8675 \times 10^{-27} kg$ for $C - H_2$ pairs). For the $H_2 - H_2$ pair, $\theta(T) = 5.4986/T$, for the $C - H_2$ pair, $\theta(T) = 2.7781/T$. To test the quantum effect, we performed GCMC to simulate the hydrogen adsorption in an empty box and plotted the calculated pressure against the set pressure (Figure 2) at 293K and 77K. It is found that the quantum effect at 293K is very small but it becomes very significant at 77K.

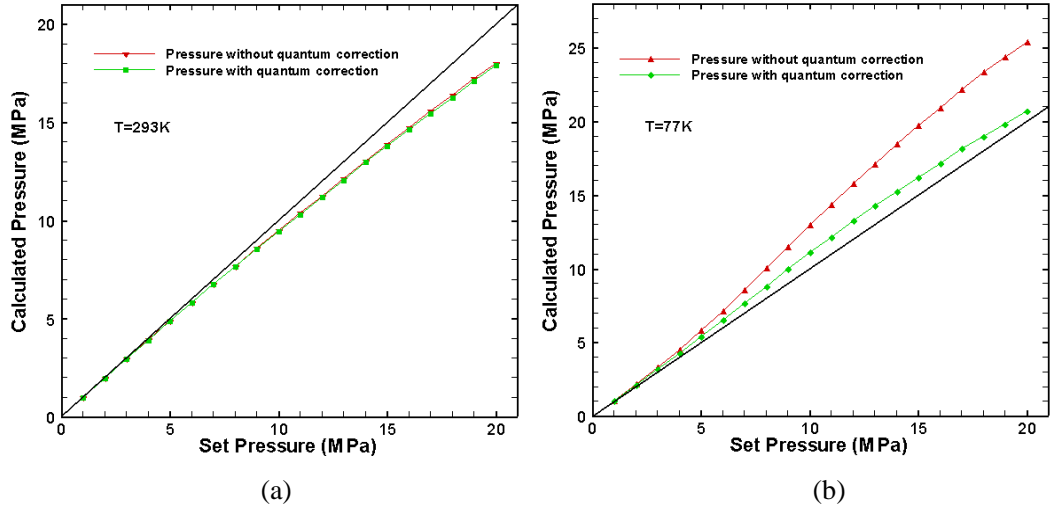

(a)                                                             (b)

Figure 2. Quantum effect on pressure at 293K and 77K (the black solid line refers to x=y(x)).

To consider the fugacity at 77K, we calculated the fugacity coefficient according to the data in Figure 2(b) by taking the ratio of the set pressure to the calculated pressure. The fugacity coefficient is fitted as a function of the set pressure using a 5$^{th}$ order polynomial as shown in eq. (7) (Figure 3). The same process was performed to obtain the fugacity coefficient function at 100K and 200K. The coefficients are listed in Table 1.

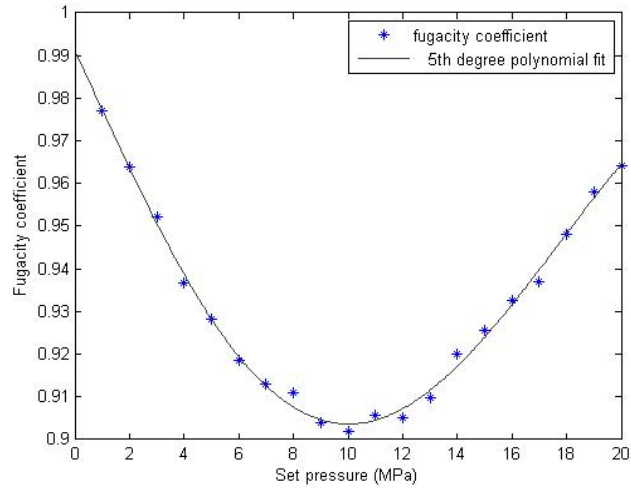

Figure 3. Fugacity coefficient as a function of set pressure.

$$\phi = \sum_{i=0}^{5} c_i P^i \qquad (7)$$

|      | $c_0$   | $c_1$                  | $c_2$                  | $c_3$                  | $c_4$                  | $c_5$                  |
|------|---------|------------------------|------------------------|------------------------|------------------------|------------------------|
| 77K  | 0.99117 | $-1.366\times10^{-2}$  | $-2.4193\times10^{-4}$ | $1.1282\times10^{-4}$  | $-4.4165\times10^{-6}$ | $4.6108\times10^{-8}$  |
| 100K | 1.0041  | $-5.9255\times10^{-3}$ | $1.1524\times10^{-3}$  | $-7.055\times10^{-5}$  | $2.6498\times10^{-6}$  | $-4.0809\times10^{-8}$ |
| 200K | 1.0071  | $3.8695\times10^{-5}$  | $1.5969\times10^{-3}$  | $-1.6331\times10^{-4}$ | $7.8198\times10^{-6}$  | $-1.4172\times10^{-7}$ |

Table 1. Parameters for equation 7.

Using the model with both fugacity and quantum corrections, we performed hydrogen adsorption in an empty box again and the calculated pressures are plotted in Figure 4. After the fugacity correction, the calculated pressure is much closer to the set pressure at 77K. As a result, after the fugacity and quantum corrections, the hydrogen gas in the reservoir modeled in the GCMC is very close to the real gas described by the LJ potential.

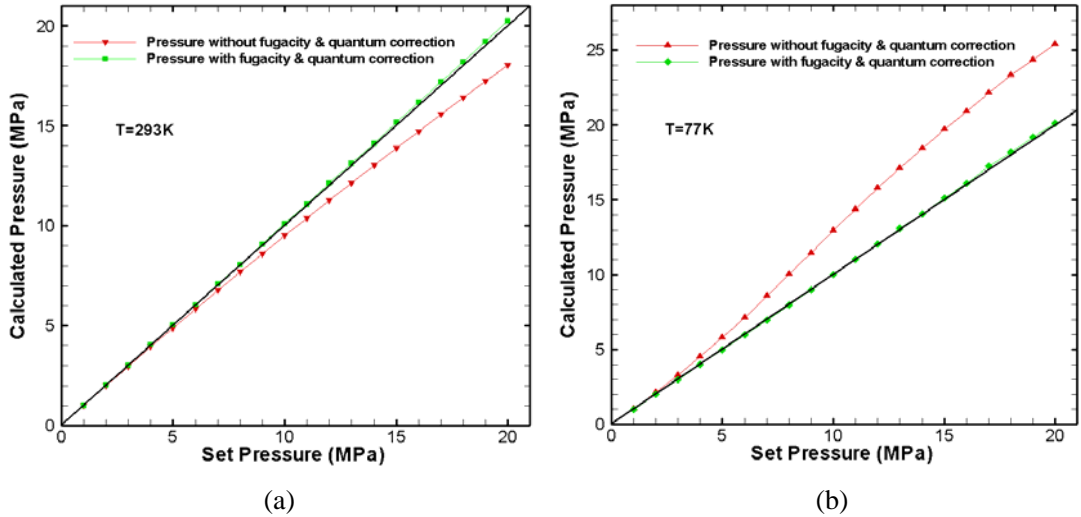

Figure 4. Calculated pressure versus set pressure after combined fugacity and quantum corrections (the black solid line refers to y(x)=x).

To investigate the combined influence of fugacity and quantum corrections on the hydrogen adsorption in carbon structures, we performed simulations on adsorption in a closely packed (10,10) CNT array at 77K and 293K with and without the corrections. The calculated adsorption wt% isotherms are presented in Figure 6. The cross section of the simulated carbon structure supercell is shown in Figure 5(c). The 2x2 supercell is expanded from the unit cell shown in Figure 5(a) in x and y directions. The supercell has a length of 19.7 Å in z-direction.

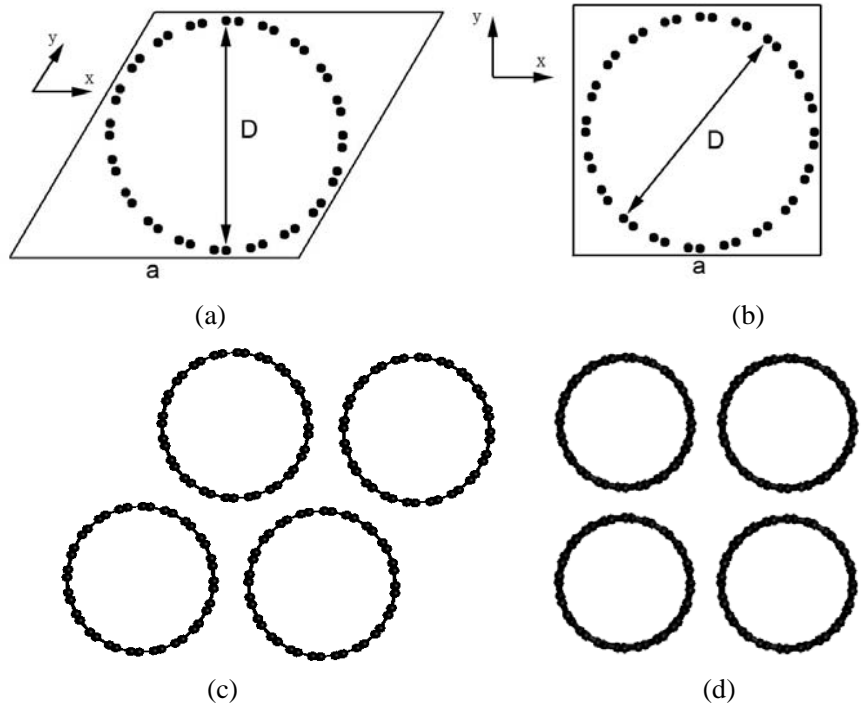

(a)　　　　　　　　　　　　　(b)

(c)　　　　　　　　　　　　　(d)

Figure 5. Cross sections of (a) a triangular unit cell, (b) a square unit cell, (c) a 2x2 triangular supercell and (d) a 2x2 square supercell

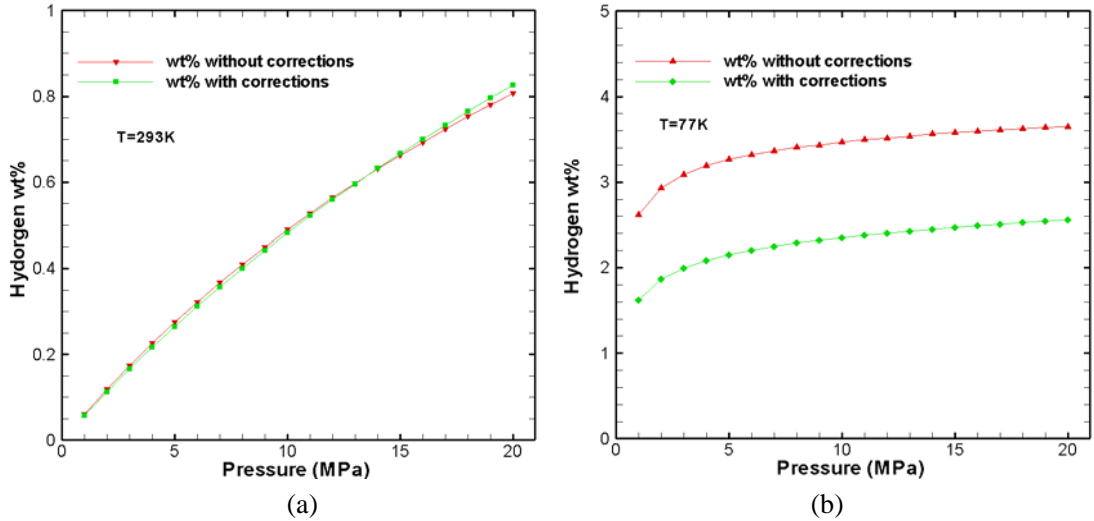

(a)　　　　　　　　　　　　　(b)

Figure 6. Combined fugacity and quantum effects on hydrogen adsorption in a closely packed (10,10) CNT array.

From Figure 6, it can be seen that at T=293K, the influence of corrections is very weak at pressures lower than 15MPa but it becomes greater as the pressure raises up to 20MPa. At temperature of 77K, the effect of the corrections is significant over the whole pressure rang. Wang et. al. [10] had similar findings on the great influence from quantum effect on the hydrogen adsorption in the interstice of CNT arrays and Patchkovskii et. al. [15] also showed the importance of the correct treatment of quantum effect on theoretical prediction on hydrogen adsorption.

## 3.2 Finite Size Effect

In this section, the finite size effects on the hydrogen adsorption in CNT arrays are studied. We performed adsorption simulations at 77K in closely packed (10,10) CNT array with different cell lengths and different cross-section areas. To change the cross-section area, the unit cell in Figure 5(a) is expanded in x- and y-directions to form 2x2 and 3x3 supercells. The calculated isotherms are plotted in Figure 7. It is found that the calculated adsorption isotherms almost overlap on one another in the 2x2(L=19.7 Å), 2x2(L=29.5 Å) and 3x3(L=19.7 Å) cases. We believe that the 2x2 supercell with a length of $L = 19.7 \text{ Å}$ is large enough to ignore the finite size effects and this supercell is used in all simulations on CNT arrays.

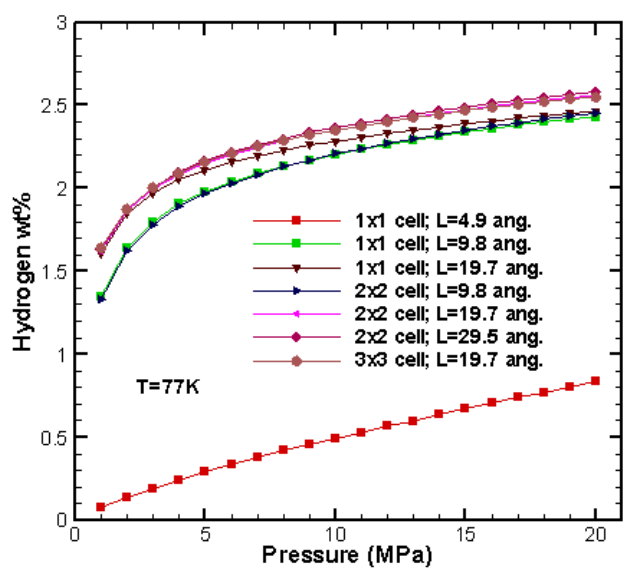

Figure 7 Finite size effect on hydrogen adsorption wt% in closely packed (10,10) CNT arrays.

## 3.3 Carbon Nanotubes
### 3.3.1 Temperature Effect

To investigate the temperature effect on the hydrogen adsorption, simulations are performed on the closely packed (10,10) CNT array at 100K and 200K. The results are presented in Figure 8 together with the results at 77K and 293K. From the figure, it is clear that the hydrogen wt% decreases as temperature increases. The wt% decreases faster with temperature at lower temperatures (77K->100K) but slower at higher temperatures (200K->293K)(e.g. a 23K temperature increase from 77K to 100K leads to 0.3wt% drop at 20MPa but a 93K temperature increase from 200K to 293K results in a 0.5wt% decrease). At low temperatures (77K->100K), the amount of wt% decrease due to temperature drop becomes smaller when the pressure increases. While at higher temperatures (200K->293K), the wt% decrease due to temperature decrease gets larger when the pressure is increased.

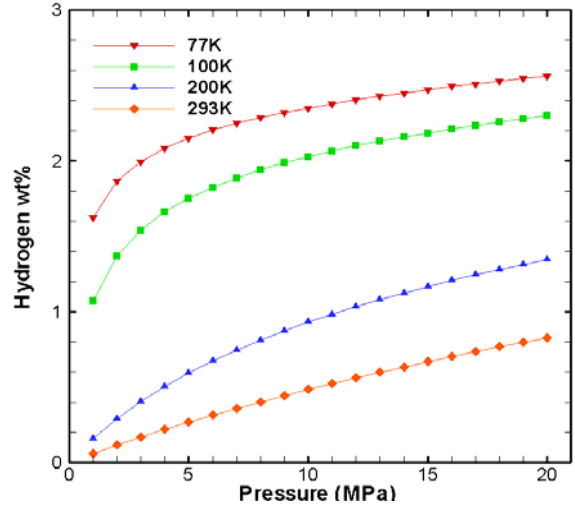

Figure 8. Temperature effect on hydrogen adsorption wt% in a closely packed (10,10) CNT array.

### 3.3.2 Diameter Effect

To study the CNT diameter effect on the hydrogen adsorption ability, triangularly packed CNT arrays with different diameters are simulated. All the arrays use the equilibrium lattice constants from ref. [29]. The van de Waals (vdW) gap, which is defined as the difference between lattice constant and the CNT diameter, is almost constant (3.1-3.2 Angstrom) among all the arrays with different CNT diameters. (6,6), (10,10), (15,15), (20,20), (26,26), (32,32) and (38,38) CNTs are studied. The relation between the diameter and the chiral indices is

$$D = \frac{\sqrt{3}}{\pi} a_{c-c} \left( m^2 + mn + n^2 \right)^{1/2} \quad (7)$$

where $D$ is the diameter, $a_{c-c}$ is the carbon bond length of a graphene which is $1.42\,\text{Å}$. $m$ and $n$ are chiral indices. Some calculated diameters are tabulated in Table 2.

| Chiral indices (m,n) | Diameter D (Å) |
|---|---|
| 6,6 | 8.14 |
| 9,9 | 12.20 |
| 10,10 | 13.56 |
| 18,18 | 24.41 |
| 20,20 | 27.12 |
| 32,32 | 43.39 |
| 38,38 | 51.53 |
| 40,40 | 54.24 |

Table 2. Diameters corresponding to CNTs with different chiral indices.

The calculated hydrogen wt% at 77K and 293K at different pressures are shown in Figure 9. It can be seen that the wt% increases with diameter increase. This is because that larger CNTs provide more space inside the tubes for hydrogen molecules. According to our calculation (Figure 9(a)), for closely packed CNT arrays to reach the DOE target of hydrogen 6.5 wt%, the (40,40) CNTs (D=54.24 Å), a high pressure of 20MPa and a low temperature of 77K are necessary. At 293K, the closely packed CNT arrays are not promising for the DOE target.

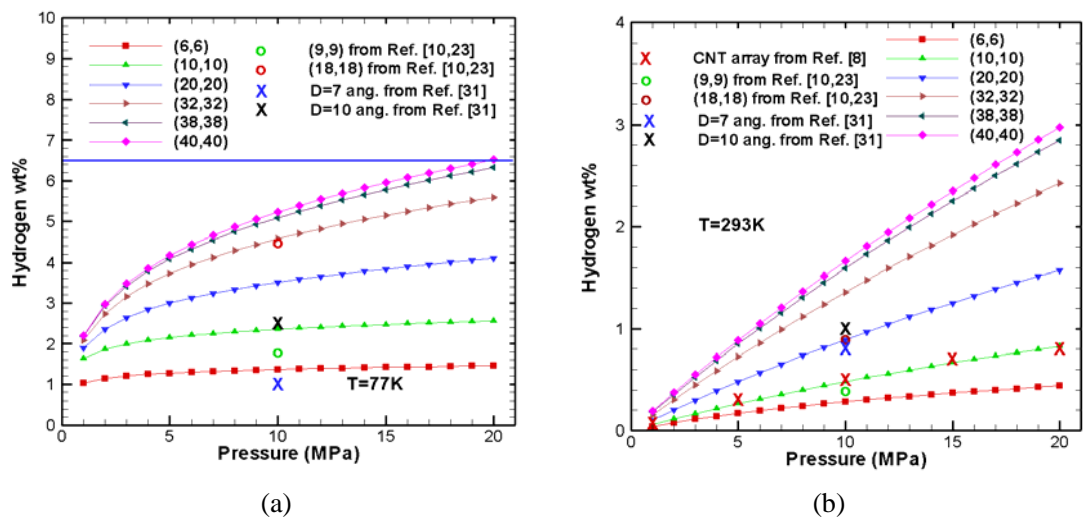

(a)                                      (b)

Figure 9. Diameter effects on hydrogen wt% at 77K and 293K in closely packed CNT arrays (the horizontal blue line in (a) refers to the DOE target of 6.5%).

Our data for the closely packed (10,10) CNT array (D=13.56) are very close to results from Levesque et. al. [8] on a D=13.3 Å, a=16.7 Å CNT array for the adsorption wt% at 293K. Their data are presented at Figure 9(b) as red crosses. It should be noted that there are three differences between our model and the model used in [8]: 1. no fugacity was considered in [8]; 2. the quadrupole-quadrupole interaction was considered in ref. [8]; 3. the quantum effect of the interaction between hydrogen molecules and carbon atoms are not considered in [8]. Our predictions for the (6,6) CNT array with D=8.14 Å is close to the data from ref. [31] at 77K for a D=7 system (blue cross in Figure 9(a)), but their data on a D=10 Å CNT array (black cross in Figure 9(a)) is even larger than our data on the (10,10) CNT array which has a D=13.56 Å. Such a difference should be the result of the quantum effect which was ignored in ref. [31]. It also significantly overestimated the wt% at 293K compared to other calculations and our data. (NOTE: the original data from ref. [8,22,23] are not expressed in wt% , the wt% data of these references are from a separate review from Meregalli et. al. [26]).

For the GCMC calculations using a different potential model (Silvera-Goldman and Crowell-Brown potential) [10,23], the results (shown as green and red circles in Figure 9(b)) are very close to our data at T=293K. At 77K, their wt% of a (9,9) CNT array (green circle in Figure 9(a)) is not very far from our data on a (10,10) CNT array. However, their data on a (18,18) array (red circle in Figure 9(a)) is much larger than our (20,20) wt%. It seems that our potential model predicts wt% data that are similar to those predicted by the model in ref. [10,23] at low hydrogen concentration, but there are large different between the data predicted by these two models when higher hydrogen concentration is involved.

### 3.3.3 CNT Spacing Effect

In this section, hydrogen adsorptions in (10,10) CNT arrays with different vdW gaps are simulated. The difference in vdW gaps are reflected by the difference in lattice constants. Both triangular unit cell (Figure 5(a)) and square unit cell (Figure 5(b)) are used. The adsorption isotherms for triangular CNT arrays at 77K and 293K are presented in Figure 10. It can be seen that the wt% increases obviously with the vdW gap increase. The amount of wt% increase due to the vdW gap increase becomes larger at higher pressure. From our calculation, at the temperature of 77K, the (10,10) CNT triangular array with a lattice constant of $26.723 \overset{o}{A}$ can reach the DOE 6.5 wt% target at a high pressure of 20MPa. Changing the vdW gap will not make (10,10) array a promising hydrogen storage media at room temperature even though the wt% increase almost linearly with the increased vdW gap. The squarely packed (10,10) CNT arrays are also simulated with the same vdW gaps as those used in the triangular arrays and the results are shown in Figure 11. The squarely packed arrays can adsorb more hydrogen than the triangular arrays for the same vdW gaps. This is because that the square arrays are more loosely packed so that the interstice spaces formed by the CNT walls are larger, and thus more hydrogen molecules are adsorbed in these spaces. For the square arrays, a lattice constant of $24.723 \overset{o}{A}$ is almost enough to reach the 6.5 wt% target at 77K and 20MPa.

Our data for the (10,10) CNT triangular array (D=13.56) with a=$20.723 \overset{o}{A}$ are close to the results from Levesque et. al. [8] on a D=$13.3 \overset{o}{A}$, a=$19.3 \overset{o}{A}$ CNT array (red crosses in Figure 10(b)), but our data are slightly smaller. As state in Section 3.3.2, the only differences between our model and the model used in ref. [8] are the consideration of fugacity, quadrupole-quadrupole interactions and $C-H_2$ quantum effect. The former two factors are proved to be not significant according the discussions in Section 2.2 and 3.1. It is also found in Section 3.1 that the quantum effect brings down the hydrogen wt%. As a result, the treatment of $C-H_2$ quantum effect should be the reason why our data is smaller than the data of ref. [8].

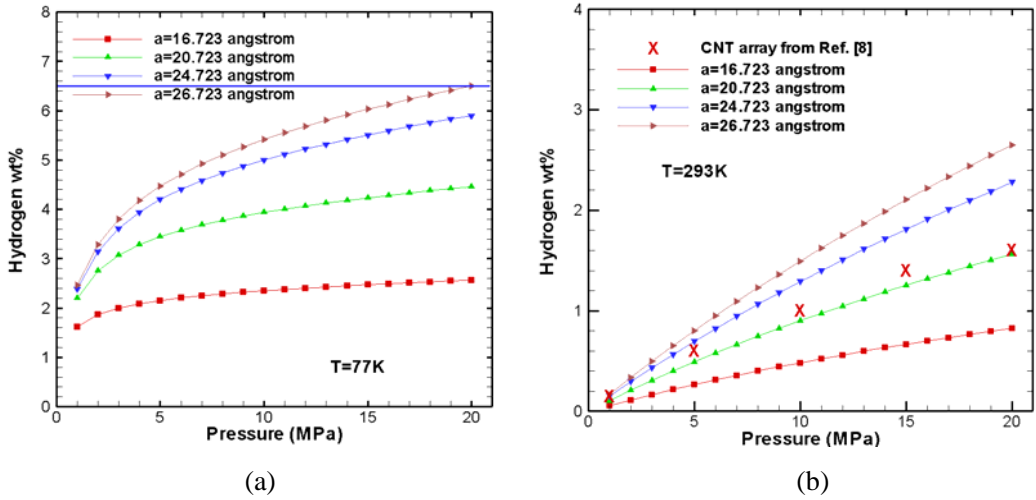

(a)                                                       (b)

Figure 10. Spacing effect on wt% for triangularly packed (10,10) CNT arrays at 77K and 293K. (the horizontal blue line refers to the DOE target of 6.5%)

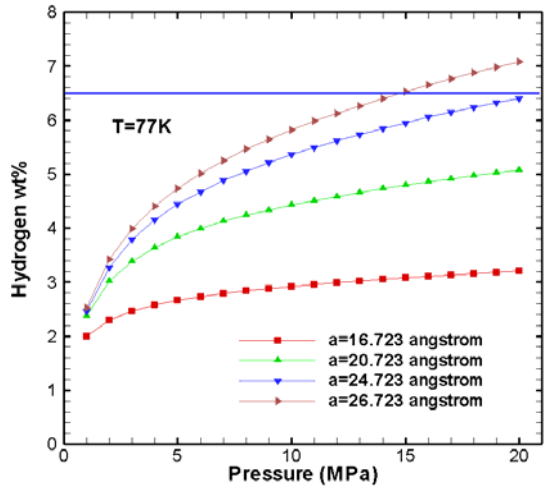

Figure 11. Spacing effect on wt% for squarely packed (10,10) CNT arrays at 77K. (the horizontal blue line refers to the DOE target of 6.5%)

**3.4 Graphite Nanofibers (GNF)**

Beside CNT arrays, graphite nanofibers (GNF) are also regarded as an alternative media for hydrogen storage [14,15,22]. In this section, the hydrogen adsorption in GNFs is simulated using GCMC. The GNFs are modeled by single graphene sheet stacks shown in Figure 12. The interlayer distances H changes to simulate the spacing effect on the hydrogen adsorption capacity.

The simulation supercell has a size of 19.7 Å in z-direction and 21.3000 Å in x-direction. The cell size in y-direction varies as the interlayer distance H changes. Since the x-, y-, z-dimensions of the supercell are not smaller than those of the smallest system to ignore the finite size effect in Section 3.2, we believe that there is no finite size effect in the calculations in this section. In this work, the $C-H_2$ potential change due to the curvature of the CNT is ignored, and thus the same potential model used in CNT system is used to model the GNF structures.

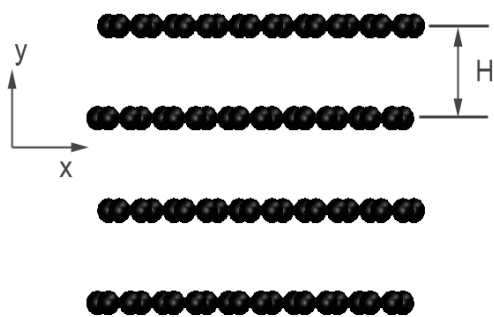

Figure 12. Simulation supercell of GNF structure.

GNFs with H of 3.414, 4.414, 5.414, 6.414, 7.414, 8.414, 9.414, 12.900 and 15.000 Å are simulated at 77K and 293K. The calculated hydrogen wt% isotherms for GNFs with H larger than 5.414 Å are plotted in Figure 14. It is found that there is no hydrogen can be adsorbed if the equilibrium H=3.414 Å or H=4.414 Å is used since the interlayer spaces are too small for hydrogen to diffuse into the structure. Sabir et. al. [9] also claimed that molecular hydrogen does not readily intercalate into pure graphite unless the distance between the graphene sheets is artificially enlarged. The calculated wt% data for GNF with different interlayer distances from calculations of ref. [10,22,31] are also plotted in Figure 14. The data from [31] (circles) for GNFs with H=7 Å and H=10 Å are much larger than our predictions for GNFs with H=7.414 Å and H=9.414 Å, respectively, at 77K. But at 293K, ref. [31] predicted hydrogen wt% for a H=10 Å GNF that is close to our data for a H=9.414 Å system. We believe the large discrepancies at low temperatures are due to that quantum effect was not considered in ref. [31]. Quantum effect greatly reduces the predicted hydrogen wt% at 77K while does not have much influence at 293K, as discussed in Section 3.1. Compared to the data from Wang et. al. [10,22] (crosses), our data on a H=6.414 Å GNF agrees well with their data for a H=6 Å GNF at both 77K and 293K, but our predictions for a H=9.414 Å system are way smaller than Wang's data for a H=9 Å system at both 77K and 293K. As discussed at the end of Section 3.3.2, Wang et. al. used a different potential model and that is likely the reason why there are differences between their predictions and our data. From our calculation, the GNF with H=15 Å is needed to reach the 6.5 wt% DOE target and this can only happen at a low temperature of 77K and a high pressure of 20MPa.

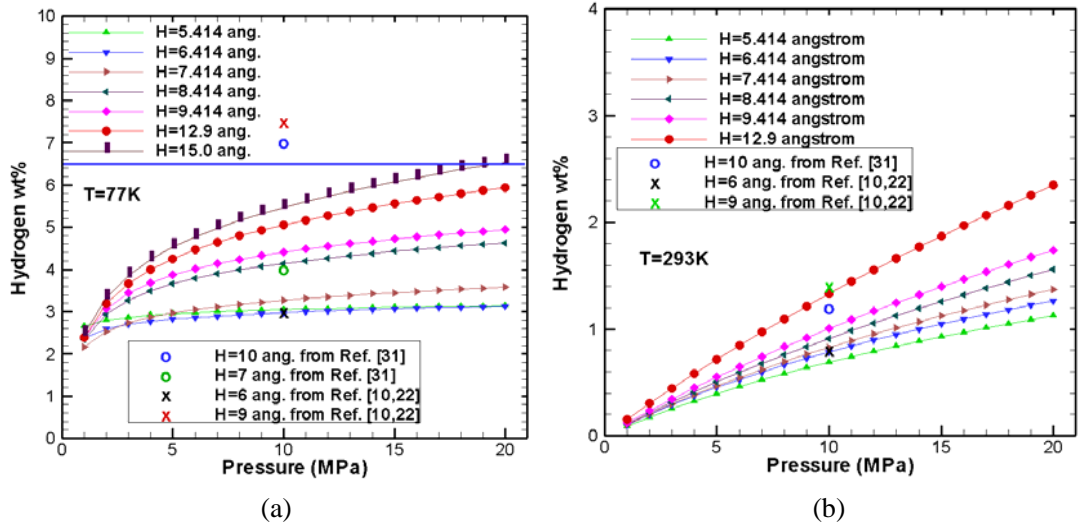

(a)                                         (b)

Figure 13. Interlayer distance effect on hydrogen adsorption wt% in GNFs. (the horizontal blue line refers to the DOE target of 6.5%)

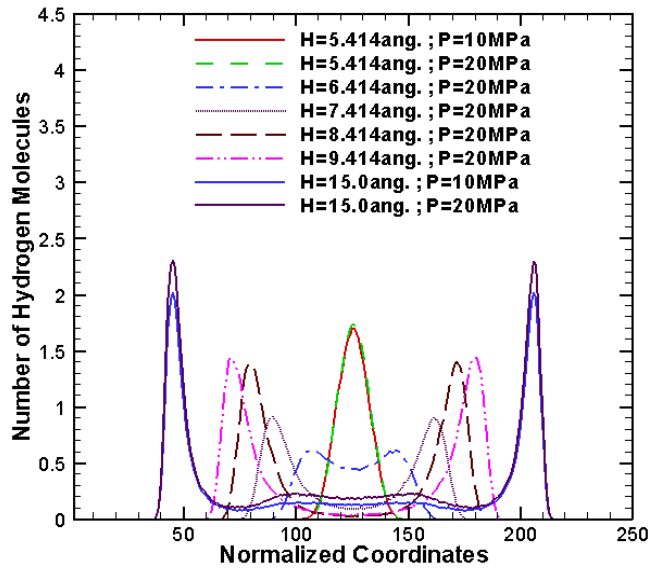

Figure 14. Hydrogen distribution between two graphene sheets with different interlayer distances.

The hydrogen equilibrium distributions inside the GNFs are calculated at 77K for different pressures and shown in Figure 14. In the figure, only the numbers of hydrogen molecules between two graphene sheets are visualized and the coordinates in y-direction are normalized against the interlayer distances. The normalized interlayer distances are divided into 250 intervals as shown in Figure 14. For the H=5.414 Å system, only one layer of molecular hydrogen can be fitted into the interlayer space and the pressures of 10MPa and 20MPa do not affect the hydrogen wt%. It is also found that for the H=4.414 Å system can not take any hydrogen at all. As a result, we believe that the interlayer distance threshold for GNFs to be able to adsorb hydrogen is between 4.414 Å and 5.414 Å. It is also found that the distribution profile for the H=6.414 Å GNF has a concave shape, meaning two layers of hydrogen molecules are being formed but the two layers still have overlaps. As a result, we believe the single layer adsorption reaches the maximum capacity when the H is between 5.414 and 6.414 Å. Patchkovskii et. al. [15] predicted a monolayer hydrogen with the maximum adsorption capacity of 3.3% can be fitted into a H=6 Å GNF. Our distribution profile suggests the same prediction and our wt% (3%) is also very close to their data. There are two layers of hydrogen molecules between two graphene sheets in GNFs with H=7.414, 8.414 and 9.414 Å. It can be seen that there are much more hydrogen molecules in each layer in the H=8.414 Å system than that in the H=7.414 Å system, and this results in a large jump of the wt% as shown in Figure 13(a). It is possible that the hydrogen in the H=7.414 Å system is still in the transition phase from one layer to two layers, meaning there are still overlaps between layers. But the H=8.414 Å system has enough space for two complete layers of molecules with no overlap. This transition process can be clearly seen from the distribution profile change from the H=5.414 Å system to the H=8.414 Å system. For the H=15 Å system, two complete layers hydrogen molecules can be adsorbed and two other layers are emerging at coordinates of 100 and 200. The hydrogen number density of these two emerging layers increases as the pressure increases from 10MPa to 20MPa.

Patchkovskii et. al. [15] also predicted that a H=9 Å system will accommodate two close-packing monolayers of hydrogen and suggested a 6.6 wt%. However, from Figure 14, the two-layer hydrogen will not reach a maximum capacity until the interlayer distance H is near 15 Å. It should be noted that ref. [15] used a free energy method with a different treatment of quantum effect. It is possible that the different treatments of quantum effect result in the difference in the predicted hydrogen wt%. However, when the GNF system can take two layers of hydrogen with the maximum capacity, the predicted wt% from our calculation (around 6%) is close to that predicted by [15].

**3.5 C60 Intercalated Graphite (CIG)**
As discussed in last section, it is found that GNF will not be able to take hydrogen unless the interlayer distance H is enlarged artificially. It is also found that the larger the H is, the more hydrogen can be adsorbed. To enlarge the H, an intercalation method using C60 was found experimentally possible by Gupta et. al. [30]. Kuc et. al [16] used a free energy method and predicted that such a C60 intercalated graphite (CIG) can easily reach the 6.5 wt% target at a temperature of 275K and a pressure around 15MPa. In this section, the GCMC is performed on the hydrogen adsorption inside the CIG structures at 77K and 293K.

The simulation supercell is presented in Figure 15 and the interlayer distance (H=12.9 Å) is taken from the experimental values from ref. [30]. The C60 are placed in a rectangular lattice with $a_x = 10.65$ Å and $a_z = 9.84$ Å (the uneven lattice constants are the results of the uneven supercell dimensions of the graphene sheet in x- and z-directions). The size of the supercell is chosen so that no finite size effect could influence the results. The results are shown in Figure 16 together with the hydrogen wt% for a GNF with the same interlayer distance. It can be seen that the CIG has largely reduced wt% values compared to the GNF. The hydrogen distributions between two graphene sheets in the CIG and GNF at 77K and 20MPa are plotted in Figure 17. It is very interesting to find out that the number of hydrogen molecules between too graphene sheets are not greatly reduced even though the C60s occupy a lot of the interlayer spaces (e.g. 219 molecules are adsorbed in the CIG and 242 molecules are adsorbed in the GNF at 77K and 20MPa). Kuc et. al. [16] suggested that the C60 lowered the interaction free energy between hydrogen and the carbon structure and compensated the effect of reduced effective adsorption space due to C60s. It is the C60s which add the weight of the carbon structure that lead to a much smaller hydrogen wt% in CIG.

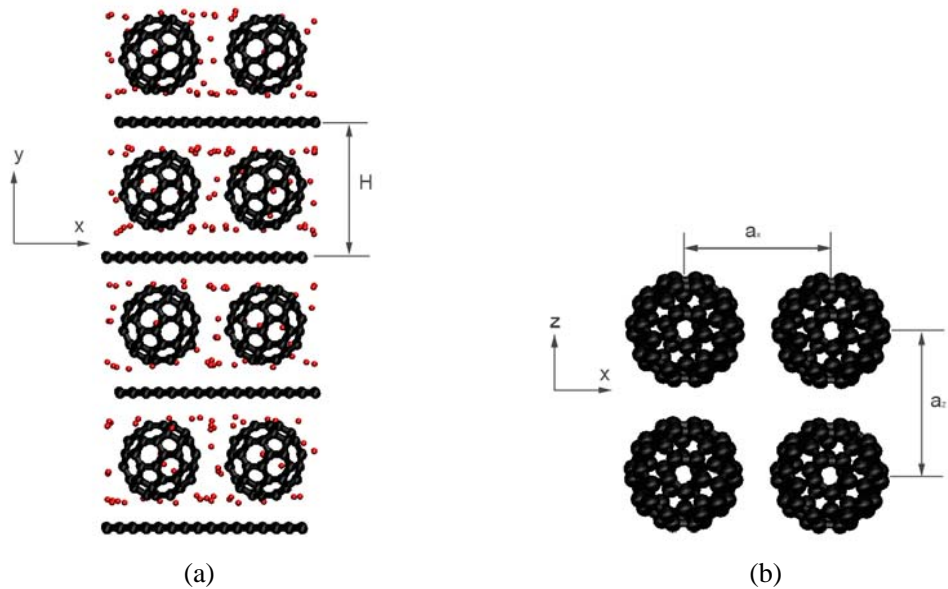

(a)          (b)

Figure 15. Simulation supercell of the CIG system: (a) projection on x-y plane, (b) projection on x-z plane without the graphene sheets.

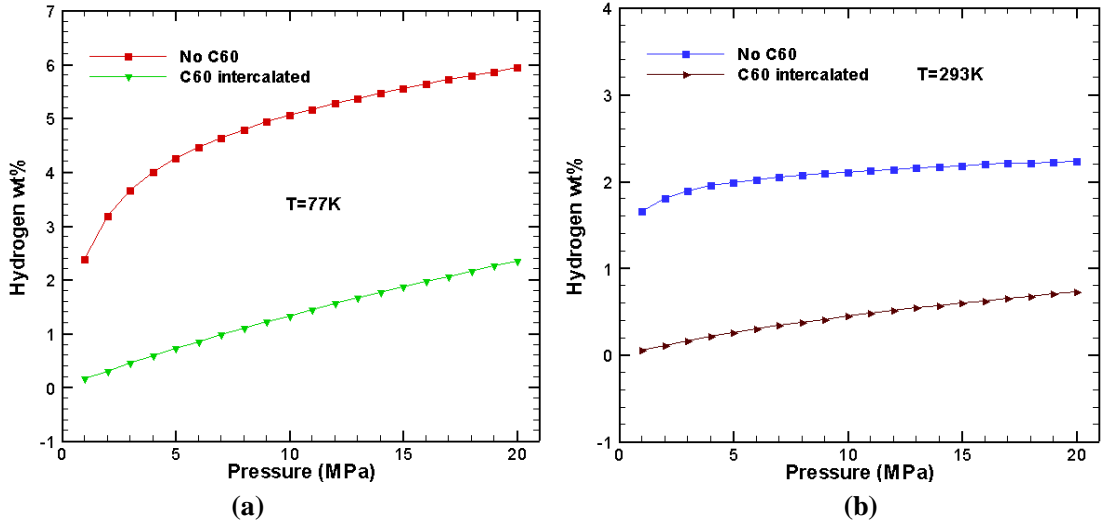

Figure 16. Comparison of hydrogen wt% in graphite with C60 intercalation and without C60 intercalation.

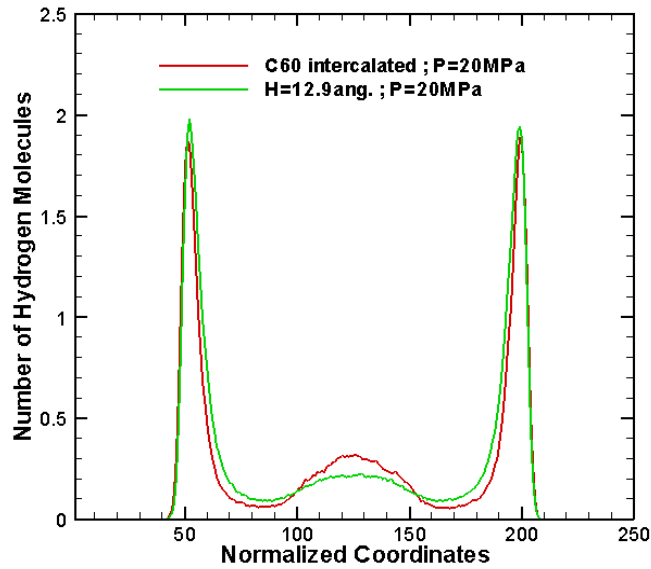

Figure 17. Hydrogen distribution between two graphene sheets with or without C60 as pillars.

The spaces inside the C60 cages do not provide reversible adsorption space for hydrogen molecules, however, we did not exclude those spaces in our calculations as did in the work by Kuc et. al. [16]. We found that each C60 cage is capable of containing only one hydrogen molecule. Kuc et. al. [16] used a free energy method together with a real gas equation of state to calculate the hydrogen physisorption wt% in the CIG and found wt% values (e.g. 300K, 20MPa, 6%) much higher than the results of the present work. We are not aware of the exact gas equation of state used for their calculation. Kuc et. al. [16] also calculated active volumes of hydrogen storage, which was defined as the space of attractive H2-host potential using a LJ potential parameterized on the basis of post-Hartree-Fock ab-initio calculation on the interaction of H2 with polyaromatic hydrocarbons. They reported an active volume of 62% for the GNF with interlayer distance of 12.5 angstrom without C60 pillars. However, using the potential model in our work, the value is only about 45%.

In any case, we believe the CIG will not be able to reach the 6.5 wt% target at room temperature and medium pressure since the upper limit is the case of the GNF with the same interlayer distance.

**4. Summary and Conclusion**

In this work, GCMC are used to study hydrogen adsorption in CNT arrays, GNF and CIG. Fugacity and quantum effects are studied. It is found that fugacity correction becomes more pronounced at high temperature (293K) and high pressure while quantum effect is significant at low temperature (77K). It is also found that temperature has great influence on the hydrogen wt%. For CNT arrays with the equilibrium vdW gaps, wt% increases with CNT diameter increase. Increase of the vdW gap will result in larger hydrogen wt% due to the increased interstitial spaces among CNTs. Squarely packed CNT arrays will hold more hydrogen than the triangularly packed arrays with the same vdW gaps. We found that GNF with equilibrium lattice constant cannot hold hydrogen at all unless the interlayer distance is enlarged artificially. The interlayer distance threshold for GNF to take hydrogen is between 4.414 and $5.414 \, \overset{o}{A}$. Hydrogen molecules form layers inside between graphene layers. The CIG hold comparative numbers of molecules to the GNF with the same interlayer distance. It is the C60s which increased the structural weight that bring down the hydrogen wt%.

Our calculated hydrogen wt% data generally agree well with other calculations at relatively low hydrogen concentration but are lower than other references which either used different potentials or ignored quantum effects. Both the different treatments of quantum effect and the different methods used to calculate the wt% have influence on the prediction at high hydrogen concentration.

From our study, CNT arrays, GNF and CIG are not promising to reach the 2010 DOE target of 6.5 wt% for hydrogen storage at room temperature and moderate pressure. To reach the target, a low temperature of 77K is necessary and large internal spaces of carbon structures are needed.

**Acknowledge**

The authors thank Dr. N. Priezjev for valuable discussions. The calculations were performed on the high performance computers of Michigan State University.